\documentclass{article}
\usepackage{spconf,amsmath, amssymb, graphicx}
\graphicspath{{images/}}
\usepackage{siunitx}
\usepackage{xspace}
\usepackage{lipsum}
\usepackage{subcaption}
\usepackage{hyperref}
\usepackage{multirow}
\usepackage{pbox}
\usepackage{tabularx}
\usepackage{xcolor}
\usepackage{verbatim}
\usepackage{pifont}
\usepackage{booktabs}
\usepackage{enumitem}
\setlist{nolistsep}

\title{Assem-VC: Realistic Voice Conversion by Assembling \\ Modern Speech Synthesis Techniques}
\name{Kang-wook Kim$^{\star \dagger}$ \qquad Seung-won Park$^{\dagger}$ \qquad Junhyeok Lee$^{\star}$ \qquad Myun-chul Joe$^{\star}$}
  
\address{$^{\star}$ MINDsLab Inc., Republic of Korea \\$^{\dagger}$ Seoul National University, Republic of Korea}

\makeatletter
\DeclareRobustCommand\onedot{\futurelet\@let@token\@onedot}
\def\@onedot{\ifx\@let@token.\else.\null\fi\xspace}

\def\eg{\emph{e.g}\onedot}

\def\etal{\emph{et al}\onedot}
\makeatother

\newcommand{\mel}{mel spectrogram\xspace}
    
\begin{document}
%
\maketitle
\begin{abstract}
Recent works on voice conversion (VC) focus on preserving the rhythm and the intonation as well as the linguistic content.
To preserve these features from the source, we decompose current non-parallel VC systems into two encoders and one decoder.
We analyze each module with several experiments and reassemble the best components to propose Assem-VC, a new state-of-the-art any-to-many non-parallel VC system.
We also examine that PPG and Cotatron features are speaker-dependent, and attempt to remove speaker identity with adversarial training.
Code and audio samples are available at \url{https://github.com/mindslab-ai/assem-vc}.
\end{abstract}
\section{Introduction}

Voice conversion (VC) is the process of converting the voice of the source speaker to the target speaker's voice without changing the linguistic information.
Early attempts in VC \cite{zhang2019joint, zhang2019improving} used parallel training data, 
where the linguistic content of the source and the target speakers are the same.
The rhythm and the intonation of converted speech needed not to be preserved.
However, recent VC studies worked on also retaining the rhythm \cite{ppg, saito2018non, lu2019one, tts_skins} or the intonation \cite{f0autovc, liu20v_transfer} of the source utterance without the need for a parallel dataset.
In other words, they only modify timbre while maintaining linguistic content, rhythm, and intonation of the original speech.
This imposes a stronger constraint than simply preserving linguistic content, and is more useful for entertainment purposes \eg{} making an expressive audiobook with celebrities' voices or providing emotional VC service.

Some prior works that adopted autoencoder-based models \cite{f0autovc, autovc, speechsplit} were able to utilize non-parallel datasets by disentangling speaker information from speech.
However, these works did not focus on maintaining the rhythm and the intonation of the original speech.
Other attempts used phonetic posteriograms (PPG) \cite{ppg, saito2018non, lu2019one}, which are suitable for capturing the linguistic content and the rhythm of speech.
Park \etal{} \cite{cotatron} proposed Cotatron, a linguistic feature extractor based on a multispeaker text-to-speech (TTS) model \cite{tacotron2}.
Valle \etal{} \cite{mellotron} introduced Mellotron, that can generate emotional and singing voice without explicit training data.
Cotatron and Mellotron have the drawback of requiring a transcript of the original speech, while Park \etal{} \cite{cotatron} reported using automatic speech recognition (ASR) transcription can solve this issue without degradation of quality.

In this paper, we decompose current VC systems, experiment with their components, and propose \emph{Assem-VC}.
Our contributions are as follows:
\begin{itemize}
    \item We categorize and analyze the components of existing non-parallel any-to-many VC systems and select the best combination.
    \item We examine that PPG and Cotatron features are not speaker-independent since such features incorporate speaking rate.
    \item We experiment with applying adversarial training to linguistic features to remove speaker dependency.
    \item On naturality and the speaker similarity, Assem-VC achieves the mean opinion scores comparable to the scores of the natural speech in many-to-many settings.
\end{itemize}

\section{Approach}
We choose the three current state-of-the-art VC systems capable of any-to-many conversion while maintaining the rhythm and the intonation of the source: PPG-VC, Cotatron-VC \cite{cotatron}, and Mellotron-VC \cite{mellotron}.
We break these models down to a linguistic encoder, an intonation encoder, and a decoder and compare each component.
We also suggest several modifications to enhance each component.
Finally, we propose Assem-VC, based on the best combination of components.

\subsection{Linguistic encoder}
\noindent\textbf{PPG Encoder.}
PPG are sequences that give the posterior probability distribution of phonemes for each time frame, and obtained from a pretrained speaker-independent ASR.
Various works \cite{ppg, saito2018non, lu2019one} employ PPG as speaker-independent linguistic features in VC.

\vspace{3pt}\noindent\textbf{Cotatron.}
Cotatron \cite{cotatron} is a linguistic encoder based on multispeaker Tacotron 2 \cite{tacotron2}.
Cotatron learns to predict the \mel from the text as well as the alignment between the \mel and the text transcript:
\begin{equation}\label{eq:tacotron2}
	\hat{M}_{1:i}, A_{i} = \mathrm{\underset{tts}{Decoder}}\left(
		\mathrm{\underset{text}{Encoder}}\left(T\right),
		M_{0:i-1}, z^{id}\right),
\end{equation}
where $ T, M, A, z^{id} $ are text, log \mel, alignment, and speaker representation, respectively.
The outputs of Cotatron are linguistic features $L$, calculated as following:
$L = \mathtt{matmul}\left(A, \mathrm{\underset{text}{Encoder}}(T)\right)$.

\vspace{3pt}\noindent\textbf{Mellotron Encoder.}
Mellotron \cite{mellotron} is a multispeaker voice synthesis model based on Tacotron 2 \cite{tacotron2} that can synthesize emotive and singing voice without explicit training data.
We view the concatenation of context vector calculated from text encoding and estimated alignments over all decoder time steps as the linguistic encoding of Mellotron.

\vspace{3pt}\noindent\textbf{Adversarial Cotatron.}
We speculate that Cotatron features and PPG are not speaker-independent because the speaker identity can be leaked through rhythm.
Nevertheless, the speaker information other than rhythm should not be embedded in the linguistic feature.
Thus, we adopt domain adversarial training \cite{ganin2016domain} to Cotatron features $L$.
We train an additional speaker classifier with a gradient reversal layer using linguistic features.

\subsection{Intonation encoder}

\noindent\textbf{Residual Encoder.}
Cotatron-VC introduces the residual encoder which encodes intonation into output residual features $R$ \cite{cotatron}.
Despite its careful design, whether the residual encoder outputs are speaker-independent is questionable.
It is also uncertain whether the residual encoder can properly encode the speech of unseen speakers.
The aforementioned risks may affect the quality of the output.

\vspace{3pt}\noindent\textbf{Normalized F0.}
Mellotron-VC uses fundamental frequency (F0) estimation algorithm, YIN \cite{yin} to encode intonation.
To make encoding more successful, we made some modifications and applied them to PPG-VC and Assem-VC.
We empirically observed that RAPT \cite{rapt} is more precise than YIN for estimate pitch contours, use RAPT for intonation encoder of Assem-VC.
After extracting the log F0 for each frame of all audios from a given speaker, we normalize them based on the mean and the standard deviation of the log F0 distribution of the speaker.
Since the speaker identity cannot be fully removed from the pitch contour with different unvoiced values across different speakers, we fill unvoiced segments with a constant value of -10 after the normalization.

\subsection{Decoder}
\noindent\textbf{Causal Decoder.}
Out of all the models we deal with, only Mellotron-VC used a causal decoder.
During the decoding step, the output is generated in an autoregressive manner by using the linguistic encoding, the intonation encoding, and the previous mel frames.
This architecture is capable of generating higher quality \mel{}s than non-causal decoders.
However, since a causal decoder is trained through teacher forcing, it may learn to cheat off of previous \mel frames, which are highly dependent on the source speaker.
Thus, speaker disentanglement may not be achieved even if the speaker-independent features are used.

\vspace{3pt}\noindent\textbf{Non-causal Decoder.}
A fully convolutional non-causal decoder is used in PPG-VC, Cotatron-VC, and Assem-VC.
The structure of the decoder is the same as Cotatron-VC \cite{cotatron} except the speaker conditioning method.
Instead of using speaker embedding for conditioning, we use an additional speaker encoder to capture the variation of target speech in the entire corpus.
Extracted speaker representation conditions the decoder via conditional batch normalization layer \cite{condbn}.

\subsection{Assem-VC}
\begin{figure}[t]
    \centering
    \includegraphics[width=\linewidth]{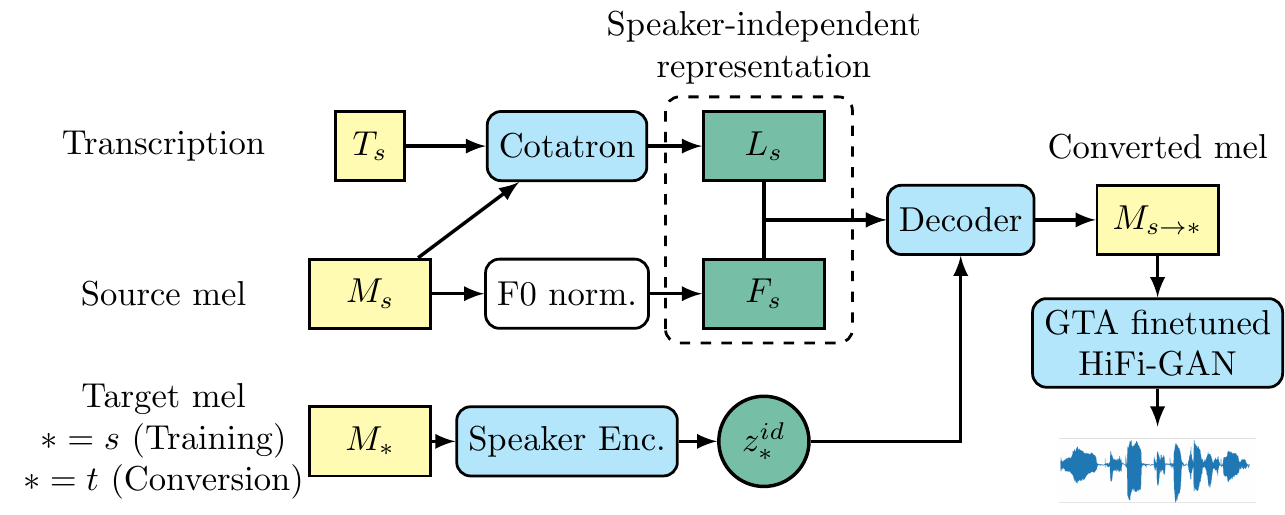}
    \vspace{-1.\baselineskip}
    \caption{%
    The overall architecture of Assem-VC.
    Speaker representation $ z^{id} $ is extracted from the speaker encoder conditions the VC decoder via conditional batch normalization layer.
    $s$ and $t$ are short for the source and the target, respectively.
    }
    \label{fig:overall}
\end{figure}

We experiment with two different versions of Assem-VC; one uses Cotatron as the linguistic encoder and the other uses adversarial Cotatron.
The linguistic features are concatenated with normalized F0 and fed to a non-causal VC decoder.
The decoder uses an additional speaker encoder for speaker conditioning.
For the vocoder, we use \textit{ground truth alignment} (GTA) finetuned HiFi-GAN.
Further details about GTA finetuning are described in \ref{training}.
Fig. \ref{fig:overall} shows the architecture of our proposed VC system, Assem-VC.
Cotatron in the figure can be replaced with adversarial Cotatron.

\section{Experiments}
\subsection{Training} \label{training}
\noindent\textbf{Dataset.}
VCTK \cite{vctk} and LibriTTS \cite{libritts} are utilized
in our experiments.
First, following the previous literature \cite{cotatron},
we split VCTK into 80\% train set, 10\% validation set, and 10\% test set, without any overlap of the transcript between them.
Note that the speakers of VCTK are shared across the splits.
Then, the systems are trained on the union of VCTK train set and LibriTTS \emph{train-clean-100} subset.
Finally, VCTK test split is used to evaluate many-to-many conversion,
while the \emph{test-clean} subset of LibriTTS is used as the source audio for the unseen speakers to evaluate any-to-many conversion.

To enhance the stability of the training process,
only the speakers with more than 5 minutes of utterances in LibriTTS \emph{train-clean-100} split are utilized.
We only used audios shorter than 10 seconds for efficient batching.
All audios are downsampled to \SI{22.05}{\kilo\hertz} and then normalized.
The \mel configuration is the same as used in Park \etal \cite{cotatron}.

\vspace{3pt}\noindent\textbf{PPG-VC.}
A pretrained wav2letter++ \cite{wav2letter++} is utilized to extract the speaker-independent PPG.
The ASR model is trained using LibriSpeech \cite{librispeech} and Common Voice \cite{comvoi}.
The CER and the WER for the LibriSpeech \textit{dev-clean} split are 2.32\% and 6.82\% without language model.
Since the frame rate of PPG is different from that of \mel, PPG are linearly interpolated to match the length of the \mel.

\vspace{3pt}\noindent\textbf{Cotatron-VC \& Assem-VC.}
For Cotatron-VC and Assem-VC, we first train Cotatron and train the whole VC system with the pretrained Cotatron fixed.
To stabilize the alignment learning of Cotatron, we initially train using only the LibriTTS \emph{train-clean-100} subset and later add VCTK.
For adversarial Cotatron, the structure and training strategy of the speaker classifier are identical to those proposed in Zhang \etal{} \cite{multilingual_tts}.
We follow the settings of Park \etal \cite{cotatron} for other training details.

\vspace{3pt}\noindent\textbf{Mellotron-VC.}
Unlike Cotatron-VC, Mellotron-VC \cite{mellotron} is trained in a single stage.
The transfer learning strategy from Cotatron-VC is employed for stable alignment learning.
Similar to Assem-VC, we normalize the pitch with respect to the speaker.
Other settings and architecture are consistent with the original Mellotron.

\vspace{3pt}\noindent\textbf{HiFi-GAN Vocoder.}
All systems employ HiFi-GAN \cite{hifigan} as the vocoder.
HiFi-GAN is pretrained with 123 hours of proprietary multi-speaker TTS dataset
and then finetuned on the train split of VCTK.

\vspace{3pt}\noindent\textbf{GTA Finetuning.}
Inspired by Shen \etal \cite{tacotron2}, we utilize GTA \mel{}s when finetuning HiFi-GAN after training the VC system.
HiFi-GAN takes the reconstructed \mel from the VC system as the input and learns to predict the source speech.
To the best of our knowledge, this is the first application of GTA finetuning for VC. 

\subsection{Conversion and evaluation}
We evaluate the many-to-many scenario by testing the conversion between the speakers from VCTK.
For the any-to-many scenario, we select a speech from \emph{test-clean} subset of LibriTTS as the unseen source speakers.
In both scenarios, only the speakers from VCTK are selected as target.

The systems are mainly evaluated with two subjective metrics
with the same details as \cite{cotatron}:
mean opinion score (MOS) and degradation mean opinion score (DMOS).
MOS measures the naturalness of speech,
and DMOS measures the speaker similarity between the target speaker's speech and the converted speech.

\subsection{Quantifying the degree of disentanglement}
To quantify the degree of speaker disentanglement of features,
we measure the accuracy of the speaker classifier trained with such features.
The speaker classification accuracy (SCA) is higher if the features are more entangled with the speaker identity.
The classifier is trained and tested on VCTK, and SCA values are reported after the validation accuracy plateaus for each feature.
The classifier consists of 4 convolutional layers and batch normalization layers, followed by a max-pooling layer and a dense layer with dropout.
Unlike other features, the size of alignments is not fixed, due to varying lengths of \mel and text.
To reduce the text index dimension, we compute weighted text indices for each frame using the alignments values to train a classifier.

\section{Results}

\subsection{Degree of disentanglement}
We measure the SCA of the features to compare their degree of disentanglement,
which is shown in Table \ref{tab:disentanglement}.
Since speaker information such as speaking rate still remains in linguistic features, SCA of PPG and Cotatron features $L$ are higher than random guessing.
Since the Cotatron features have SCA close to that of the PPG, we speculate that the PPG and the Cotatron features are equally capable in terms of speaker disentanglement.
Lastly, the normalized pitch contour $F$ is more speaker-independent than the residual encoding $R$ from Cotatron-VC.
\begin{table}[h]
	\caption{%
		Degree of speaker disentanglement.
		Rand. denotes the case of the classifier trained with the features filled with random numbers.
		$adv$ indicates the features from adversarial Cotatron.
		$L$, $F$, $R$, $A$, $M$ denote linguistic features, normalized F0, residual features, alignments, \mel.
	}
	\label{tab:disentanglement}
		\vspace{-0.5\baselineskip}
	\centering
	\resizebox{0.95\linewidth}{!}{%
	\begin{tabularx}{\linewidth}{Xccccc}
		\toprule
		\textbf{Feature} & Rand. & PPG & $ L $ & $ (L, F) $ & $ (L, R) $ \\
		\midrule
		\textbf{SCA} & 0.9\% & 31.6\% & 32.8\% & 41.7\% & 57.9\% \\
		\toprule
		\textbf{Feature} & $ A $ & $ A_{adv} $ & $ L_{adv} $ & $M$ \\
		\midrule
		\textbf{SCA} & 35.8\% & 24.0\% & 33.3\% & 99.5\% \\
		\bottomrule

	\end{tabularx}}
\end{table}

\begin{table*}[t]
	\centering
	\setlength\tabcolsep{4.5pt}
	\caption{%
	    Comparison of components and performances of the examined VC approaches.
	    \textsuperscript{*} \emph{GTA} denotes whether the HiFi-GAN vocoder is finetuned with a ground-truth aligned \mel.
	    Adv. Cotatron is short for adversarial Cotatron.
	   }
	   \vspace{-0.5\baselineskip}
	\label{tab:results}
	\resizebox{1.\linewidth}{!}{%
	\begin{tabularx}{\textwidth}{Xcccccccc}
		\toprule
		\multirow{2}{*}{\textbf{Approach}} & \multicolumn{4}{c}{\textbf{Components}} & \multicolumn{2}{c}{\textbf{Many-to-many}} & \multicolumn{2}{c}{\textbf{Any-to-many}} \\
		\cmidrule{2-9}
		& Linguistic & Intonation & Decoder & GTA \textsuperscript{*}  & MOS & DMOS & MOS & DMOS \\
		\midrule
		\multicolumn{5}{l}{Source as target} & $ 4.14 \pm 0.08 $ & $ 1.71 \pm 0.14 $ & -- & -- \\
		\multicolumn{5}{l}{Target as target} & $ 4.14 \pm 0.08 $ & $ 3.94 \pm 0.12 $ & -- & -- \\
		\midrule
		\multirow{2}{*}{PPG-VC} & \multirow{2}{*}{PPG} & \multirow{2}{*}{F0 (RAPT)} & \multirow{2}{*}{Non-causal} & \ding{55} & $ 2.47 \pm 0.09 $ & $ 2.99 \pm 0.13 $ & -- & -- \\
		& & & & \ding{51} & $ 3.81 \pm 0.09 $ & $ 3.78 \pm 0.13 $ & $ \mathbf{3.63 \pm 0.11} $ & $ 3.66 \pm 0.13 $ \\
		\multirow{2}{*}{Cotatron-VC} & \multirow{2}{*}{Cotatron} & \multirow{2}{*}{Residual} & \multirow{2}{*}{Non-causal} & \ding{55} & $ 2.60 \pm 0.12 $ & $ 3.44 \pm 0.15 $ & -- & -- \\
		& & & & \ding{51} & $ 3.75 \pm 0.09 $ & $ 3.79 \pm 0.13 $ & $ 2.99 \pm 0.15 $ & $ 3.31 \pm 0.15 $ \\
		\multirow{2}{*}{Mellotron-VC} & \multirow{2}{*}{Tacotron2} & \multirow{2}{*}{F0 (YIN)} & \multirow{2}{*}{Causal} & \ding{55} & $ 2.94 \pm 0.11 $ & $ 2.87 \pm 0.14 $ & -- & -- \\
		& & & & \ding{51} & $ 3.36 \pm 0.10 $ & $ 3.15 \pm 0.13 $ & $ 3.42\pm 0.11 $ & $ 2.84 \pm 0.13 $ \\
		\midrule
		\multirow{4}{*}{Assem-VC} & \multirow{2}{*}{Cotatron} & \multirow{2}{*}{F0 (RAPT)} & \multirow{2}{*}{Non-causal} & \ding{55} & $ 2.87 \pm 0.11 $ & $ 3.27 \pm 0.11 $ & -- & -- \\
		& & & & \ding{51} & $ \mathbf{3.91 \pm 0.10} $ & $ \mathbf{3.86 \pm 0.12} $ & $ \mathbf{3.61 \pm 0.12} $ & $ \mathbf{3.86 \pm 0.11} $ \\
		 & \multirow{2}{*}{Adv. Cotatron} & \multirow{2}{*}{F0 (RAPT)} & \multirow{2}{*}{Non-causal} & \ding{55} & $ 2.35 \pm 0.13 $ & $ 3.24 \pm 0.14 $ & -- & -- \\
		& & & & \ding{51} & $ 3.56 \pm 0.11 $ & $ 3.69 \pm 0.13 $ & $ 3.25 \pm 0.13 $ & $ 3.48 \pm 0.14 $ \\
		\bottomrule
	\end{tabularx}}
	\setlength\tabcolsep{6pt}
\end{table*}

To reveal the effect of adversarial training, we additionally measure the SCA of linguistic features and alignments of adversarial Cotatron.
Adversarial training helped to reduce the speaker dependency of alignments $A$, but it does not affect the dependency of Cotatron features $L$.
We suspect that phoneme-wise speaking rates are too tightly coupled with speaker identity to be disentangled.
Also, adversarial training hinders Cotatron from learning alignments.
As shown in Table \ref{tab:results}, there is a degradation of quality when adopting adversarial training on Cotatron.

\vspace{-0.5\baselineskip}
\subsection{Many-to-many conversion}
As presented in Table \ref{tab:results},
our proposed combination Assem-VC outperforms all other methods in terms of both MOS with DMOS,
and both of its scores close to those of natural speech.
GTA finetuning plays a key role in removing noisy artifacts, improving MOS and DMOS of all models.
We observe that the increase in MOS and DMOS after GTA finetuning is relatively low for Mellotron-VC.
We suspect that causal decoders have an advantage in the amount of information available during generation of each mel frame.
However, such setting also has a negative impact on speaker disentanglement, resulting in the lowest DMOS across the board.
PPG-VC has the closest MOS to Assem-VC, but it still performs poorly on DMOS.
Cotatron-VC shows worse performance than Assem-VC on both MOS and DMOS, with or without the GTA finetuning.
These results suggest that a combination of Cotatron and F0 estimator
is better than PPG encoder and the residual encoder
as linguistic and intonation encoder from the source speech.

\vspace{-0.5\baselineskip}
\subsection{Any-to-many conversion}
To evaluate the generalization to the unseen speakers, we also explore any-to-many situation, namely to convert the speech of an arbitrary speaker to the voices of multiple speakers.
As shown in Table \ref{tab:results}, the overall results of the any-to-many conversion are lower than the many-to-many, since any-to-many is inherently a more challenging task.
Assem-VC using Cotatron outperforms other baseline models.
PPG-VC yields a competitive MOS with Assem-VC, but a worse DMOS.
The residual encoder cannot properly encode the intonation with unseen speakers,
which causes MOS of Cotatron-VC to be the lowest.
Mellotron-VC's causal decoder fails to preserve the speaker's characteristics, resulting in the worst DMOS.

\vspace{-0.5\baselineskip}
\section{Conclusion}
In this paper, we compared the existing methods on any-to-many non-parallel VC.
We decompose these models into three components: a linguistic encoder, an intonation encoder, and a decoder.
We plug different combinations of modules into these components and report the results.
First, our results suggest that the Cotatron features are better than PPG
at representing the linguistic features of a speech without sacrificing the performance on speaker similarity.
Also, the traditional F0 estimator is better than the residual encoder
at disentangling the intonation from a speech.
We also show that the causal decoder has a negative effect on speaker disentanglement.
For the first time, to the best of our knowledge, we apply GTA finetuning for VC, so that non-causal decoders generate a natural speech that captures the target speaker identity.
We observed that adversarial training for speaker disentanglement interferes with Cotatron's ability to learn alignments.
We select the best features and reassemble to present an any-to-many VC system, Assem-VC.
Assem-VC achieves the level of natural speech recordings in both MOS and DMOS for any-to-many VC with 108 target speakers.
Our results indicate that Assem-VC can be applied to real-world applications which require voice conversion indistinguishable from human speech.

\vspace{-0.5\baselineskip}
\section{Acknowledgements}
The authors would like to thank
Sang Hoon Woo, Seungu Han, and Dongho Choi from MINDsLab Inc.,
Yoonhyung Lee from Seoul National University
for providing beneficial 
feedback on the initial version of this paper.

\bibliographystyle{IEEEbib_abbrev}
\bibliography{mybib}

\end{document}